\documentclass[pra,twocolumn,superscriptaddress,showpacs]{revtex4}

\pdfoutput=1

\usepackage{amsmath,amsfonts,amssymb,bm}
\usepackage{dcolumn}
\usepackage[final]{graphicx}
\usepackage{bm}
\usepackage{comment}

\includecomment{pdffigure}
\excludecomment{pdffigure}

\begin{document}

\bibliographystyle{apsrev}

\title{Optimizing number squeezing when splitting a mesoscopic condensate}

\author{Julian Grond}
\affiliation{Institut f\"ur Physik,
  Karl--Franzens--Universit\"at Graz, Universit\"atsplatz 5,
  8010 Graz, Austria}
\author{J\"org Schmiedmayer}
\affiliation{Atominstitut der \"Osterreichischen Universit\"aten,
  TU--Wien, Stadionallee 2, 1020 Wien, Austria}
  
\author{Ulrich Hohenester}
\affiliation{Institut f\"ur Physik,
  Karl--Franzens--Universit\"at Graz, Universit\"atsplatz 5,
  8010 Graz, Austria}

\date{\today}

\begin{abstract}

We optimize number squeezing when splitting a mesoscopic Bose
Einstein condensate. Applying optimal control theory to a
realistic description of the condensate allowed us to identify a
form of the splitting ramp which drastically outperforms the
adiabatic splitting. The results can be interpreted in terms of a
generic two-mode model mapped onto a parametric harmonic
oscillator. This optimal route to squeezing paves the way to a much longer phase coherence and atom
interferometry close to the Heisenberg limit.

\end{abstract}



\maketitle


Confined atom interferometers using Bose-Einstein condensates
(BECs) offer new prospects for matter wave interferometry
\cite{cronin:07} and precision measurements. Optical dipole traps
\cite{grimm:00}, atom chips \cite{folman:02}, and
radio-frequency (rf) potentials
\cite{lesanovsky:06,hofferberth:06} provide powerful tools which
enable coherent manipulation and interference as demonstrated in a
series of recent experiments
\cite{schumm:05,hofferberth:07,jo:07,hofferberth:08,albiez:05,esteve:08}.

Atom interferometers based on BECs usually suffer from the
nonlinearity originating from atom-atom interactions, which leads
to phase diffusion \cite{javanainen:97}. A possible way out is to
seek for narrow number distributions of the split condensate--
i.e., squeezed states, which are very powerful in precision
measurements \cite{giovannetti:04,wineland:94,sch:08,appel:08}. This can be
achieved by adiabatic splitting, where the nonlinear interaction
favors narrow number distributions
\cite{javanainen:99,jo:07,esteve:08}. The disadvantage of this
scheme is the long time needed for the splitting process, within
which technical noise and additional phase diffusion might
threaten the interferometer performance \cite{burkov:07,li:08,khodorkovsky:08}.

In this paper, we show that splitting protocols, based on optimal
control theory (OCT), allow efficient number squeezing on a much
shorter time scale and drastically outperform adiabatic splitting. 
We first investigate the OCT problem in the framework of a simple
two-mode model, leading us to an intuitive interpretation of the
control strategy. The predictions of the simple model are verified for a realistic
experimental setting by applying OCT
to the many-body problem, within the framework of multiconfigurational time-dependent Hartree equations for bosons [MCTDHB(2)]
\cite{alon:08}. The fringe visibility is significantly
enhanced in case of optimized splitting, which renders this scheme
ideal for atom interferometry.

\begin{figure}[b]
  \includegraphics[width = 1\columnwidth]{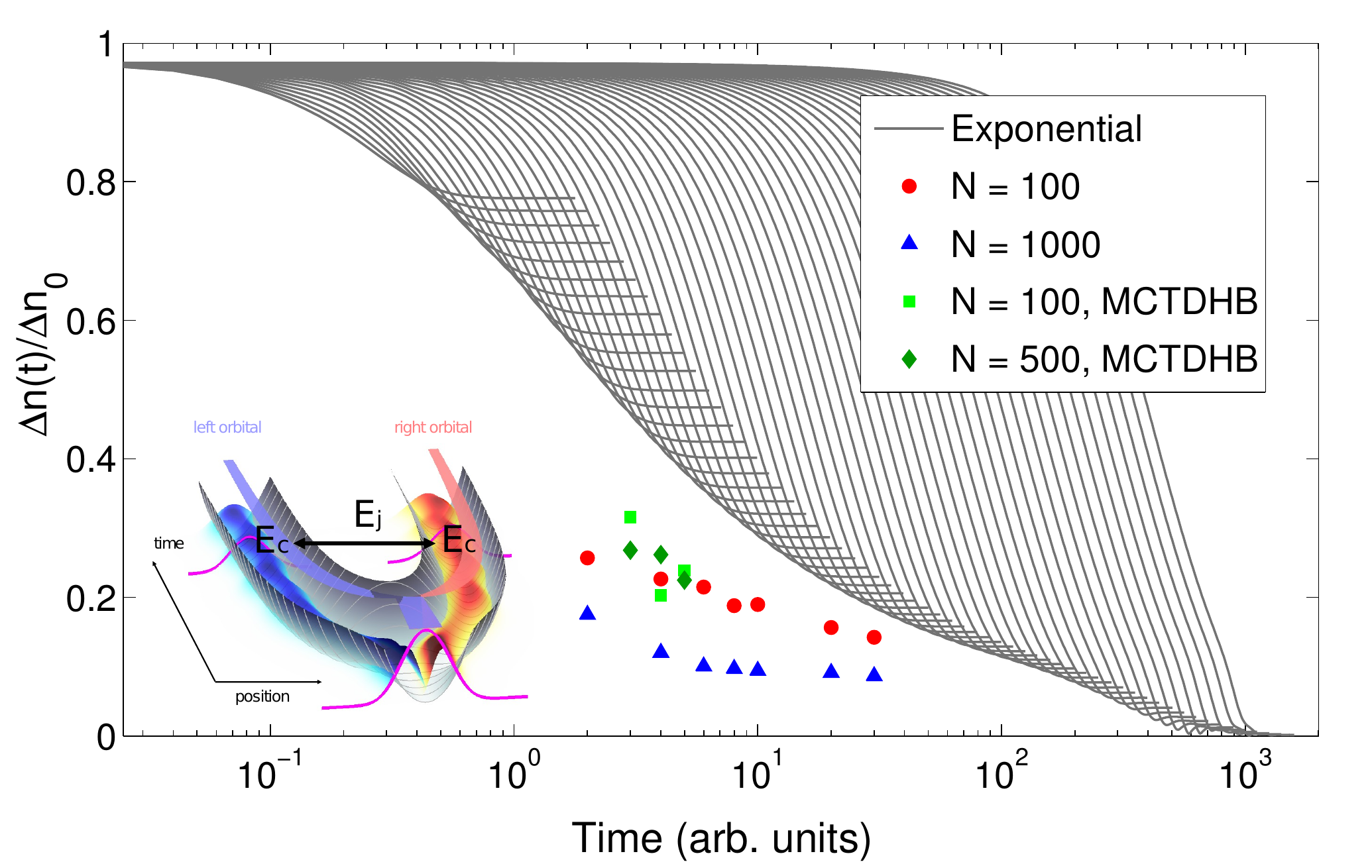}
  \caption{(Color online) Time evolution of atom number fluctuations
  for $E_j(t)=E_j(0)\,\exp[-\frac t\tau]$ and for different
  decay constants $\tau$, as computed within the generic two-mode
  model \cite{milburn:97,javanainen:99}.
  With increasing $\tau$, the splitting process is more adiabatic
  and the fluctuations in the final state are lowered. The symbols
  report results of our OCT calculations for different numbers of
  atoms and splitting times $T$, and show that OCT can significantly outperform the more
  intuitive quasi-adiabatic scheme for the exponential turning off. 
  The square (diamond) symbols correspond to solutions obtained within MCTDHB
  for $N=100$ ($N=500$) (see Fig.~3). In case of MCTDHB we take the $E_c$ value after splitting for comparison. Inset: Schematics
  of the BEC splitting process. By ramping up a double-well
  potential, the initial ground-state wave function of a single well
  becomes nonadiabatically split into two parts, denoted as left and right orbitals. 
  }
\end{figure}

Splitting a Bose-Einstein condensate is achieved by changing the confinement
potential smoothly from a single well to a double well, as
schematically shown in the inset of Fig.~1. We will assume that
the condensate wave function is modified only along a single
spatial direction $x$. To describe properly the fragmentation of
the BEC into two spatially separated condensates, we need at least
two wave functions $\phi_{L,R}(x)$, which we will refer to as left
and right {\em orbitals},\/ together with an additional part that
describes how the atoms are distributed among these two orbitals.
Close to the splitting point, where the two orbitals become
spatially separated, the system can be approximately described by
a generic two-mode model, characterized by the Hamiltonian
\cite{milburn:97,javanainen:99}
\begin{equation}\label{eq:hamtwomode}
  \hat H=-\frac 2 N E_j\,\hat J_x+\frac 1 2 E_c\,\hat J_z^2\,.
\end{equation}
Here, the Josephson energy $E_j$, which is proportional to the energy overlap of the orbitals,
accounts for tunneling whereas the charging energy $E_c$ accounts for the nonlinear
coupling of the atoms.  $\hat J_x$ and $\hat J_z$ are pseudospin
operators associated with these couplings
\cite{milburn:97,javanainen:99}: $\hat J_x$ promotes an atom from
the left to the right well, or vice versa, and $\hat J_z$ measures
the atom number difference between the two wells.

In experiments, $E_j$ is controlled indirectly by variation of
the confinement potential.  Here we use $E_j$ itself as a
control parameter in order to grasp the essential features
of the control strategy, and will lift this assumption later.

When tunneling dominates over the nonlinear interaction,
$E_j\gg E_c N$, all atoms reside in the bonding orbital
$\phi_g=\frac 1{\sqrt 2}(\phi_L+\phi_R)$, resulting in a binomial
atom number distribution with fluctuation $\Delta n_0$. When the
tunnel coupling is reduced, the nonlinear coupling favours
localization of the atoms in one of the wells. The state of lowest
energy is a superposition of different atom number states with
smaller than binomial number fluctuations. In the limit of very
small tunnel coupling, $E_j \rightarrow 0$, the ground state is such
that half of the atoms reside in the left well and the other half
in the right well, and there are no atom number fluctuations,
$\Delta n=0$.

To split the condensate, one starts from a state with
$E_j\gg E_c N$ and then turns off the tunnel coupling. A
quasiadiabatic splitting corresponds to an exponential decrease
of $E_j$, as shown in Fig.~1 for different decay constants
$\tau$ and 100-1000 atoms, corresponding to realistic experimental conditions. One observes that for a slowly varying $E_j$
the system evolves almost adiabatically and finally ends up in a
state with small atom number fluctuations. For faster splitting, the system can no
longer follow adiabatically and becomes frozen in a state with
substantially larger number fluctuations (less number squeezing).
By increasing $\tau$ by a factor of 10,
the number fluctuations in the final state drop by a factor of
approximately 2. {\em Thus, efficient atom number squeezing
comes at the price of very slow splitting.}

To improve this we are seeking for an optimal time variation of
$E_j(t)$ that brings the system to a number squeezed state with
reduced atom number fluctuations in much shorter time.
To this end, we employ OCT
\cite{peirce:88,borzi.pra:02,hohenester.pra:07} with the goal to
minimize the atom number fluctuations $(\Delta n)^2$
\begin{equation}\label{eq:cost}
  \mathcal{J}=\left(\Delta n\right)^2=
  \left< \hat J_z^2 \right>-\left< \hat J_z \right>^2\,.
\end{equation}
in the state at the final time $T$.  Within the framework of OCT,
$\mathcal{J}$ is called the {\em cost function},\/ which is
minimized under the constraint that the system's time evolution is
governed by the Schr\"odinger equation. This is done by using
Lagrange multipliers to turn the constrained minimization problem
into an unconstrained one, as discussed in some length in
Refs.~\cite{borzi.pra:02,hohenester.pra:07}.

Our OCT calculations within the two-mode model are summarized by
the circular (for N=100) and triangular (for N=1000) symbols in Fig.~1.  One
clearly sees that the optimization of number squeezing works
within a wide range of splitting times, and $E_j(t)$ sequences
found in OCT perform approximately one order of magnitude faster
in comparison to the standard exponential one.

Fig.~2 shows details to one of the $E_j(t)$ time sequences of
our squeezing optimization for N=100. OCT comes up with an
oscillating tunnel control, which leads to a drastic reduction
of the number fluctuations in comparison with the exponential
turning-off.

\begin{figure}
  \centerline{\includegraphics[width=0.6\columnwidth]{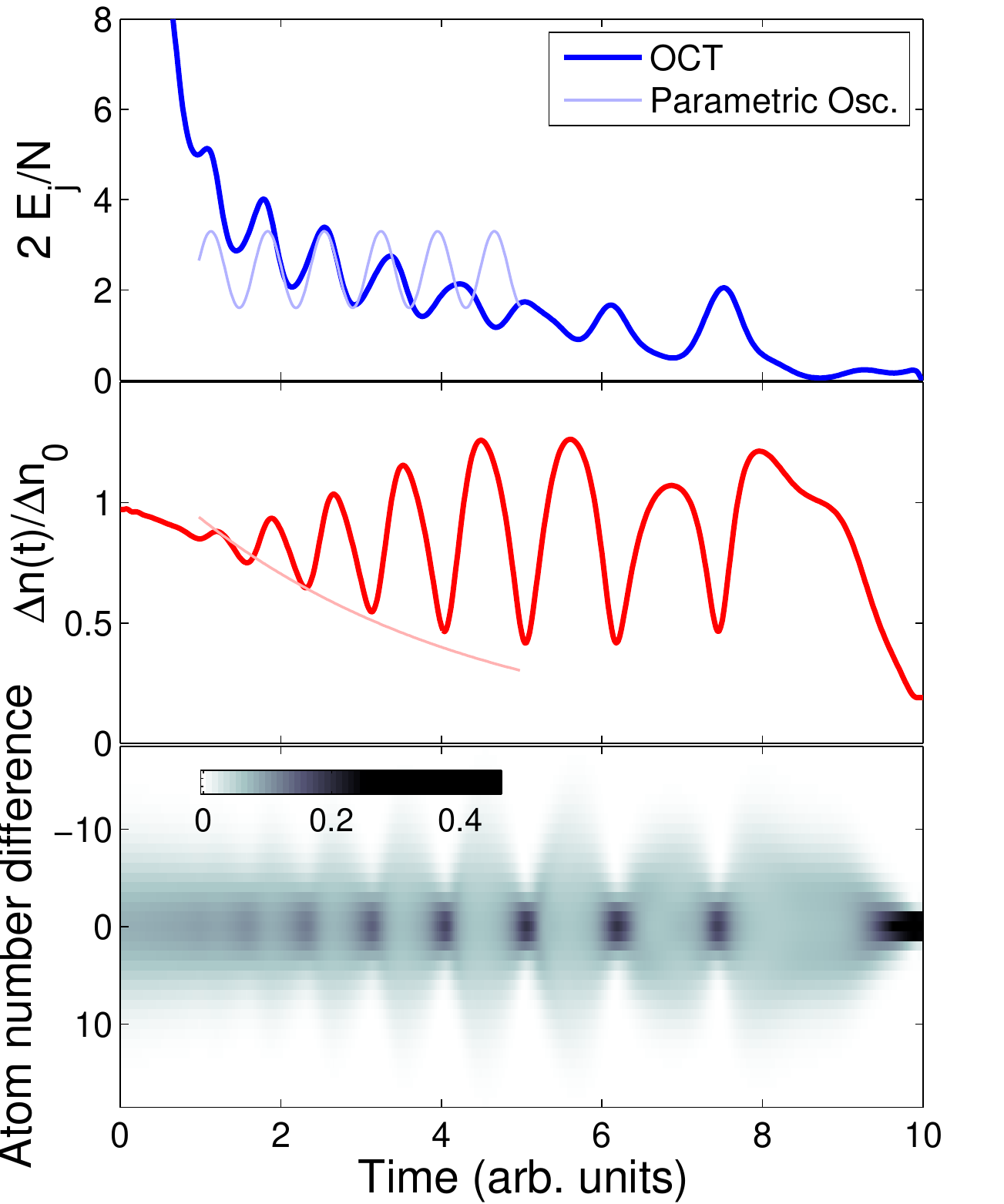}}
  \caption{ (Color online) Results of OCT calculations for the
  generic two-mode model and for an atom number $N=100$. In (upper panel), we
  plot the tunneling control $E_j(t)$, in (middle panel) the time evolution
  of the atom number fluctuations, and in (lower panel) a density plot of the
  absolute square modulus of the atom number wave function. The bright
  lines show estimates based on the parametric oscillator of
  Eq.~\eqref{eq:twomodeosc2}. The time interval has been chosen somewhat larger as in Fig.~3 to make the oscillating control mechanism more visible.  }
\end{figure}


We next investigate why the oscillating OCT tunnel coupling
drastically outperforms the more intuitive exponential decay. We show how to qualitatively understand the mechanism of the control.

For large atom numbers $N$, the time evolution of the generic two-mode
model can be approximately described by a harmonic oscillator
\cite{javanainen:99}
\begin{equation}\label{eq:twomodeosc}
  i\dot C(k)=\left[-\frac{E_j}{2}\frac{\partial^2}{\partial k^2}+
  \left(\frac{2 E_j}{N^2}+\frac{E_c}{2}\right)k^2\right]C(k)\,,
\end{equation}
where $C(k)$ is the atom-number wave function and $k$ is the number
difference between the left and right well, which is treated as a continuous variable. Introducing the
annihilation and creation operators $\hat a$ and $\hat
a^\dagger$ for the harmonic oscillator \cite{messiah:65}, we can
cast the Hamiltonian of Eq.~\eqref{eq:twomodeosc} in the form
\begin{equation}\label{eq:twomodeosc2}
  \hat H=2\tilde{E}_j/N\left[\left(\hat a^\dagger\hat a+\frac 12\right)+
  \frac{E_c N^2}{16\tilde{E}_j}\left(\hat a^2+{\hat a^{\dagger\,2}}\right)\right]\;.
\end{equation}
Here  $\tilde{E}_j/N=E_j/N+E_c N/8$ is the renormalized oscillator frequency. Equation~\eqref{eq:twomodeosc2} is the Hamiltonian for a parametric harmonic oscillator \cite{scully:97} with resonance frequency $\omega_{res}\approx 2E_j/N+E_c N/4$.

When the oscillator is driven
with approximately twice the resonance frequency, 
which is approximately the period of the OCT control in Fig. 2,
the width of the
initial ground state wavepacket starts to oscillate and becomes
strongly squeezed. The predictions of the parametric
oscillator model are plotted in Fig. 2 for the oscillating control field
indicated in the upper panel: indeed, as can be observed from the number
fluctuations in the middle panel, the envelope of $\Delta n(t)$ decreases
in a fashion similar to the results of the OCT calculations.
Thus, if we turn off $E_j$ at the lower turning point of $\Delta n(T)$, we
freeze the system in a squeezed number state [although
the detailed freezing sequence of $E_j(t)$ happens to be more complex
in case of OCT].


We now address the question whether our findings would prevail in case of a more realistic
modelling of the many-body splitting process. Contrary to the
two-mode model, where all details of the condensate orbitals are
embodied in $E_j$ and $E_c$, in general the control of
$E_j$ and $E_c$ is indirect \emph{through} the condensate
orbitals, which, in turn, can be manipulated by means of the
confinement potential $V_\lambda(x)$. Here, $\lambda(t)$ is a
control parameter that describes the variation of the confinement
potential when changing the external parameters, such as currents
through the microtrap wires or frequency and strength of
additional rf fields \cite{lesanovsky:06}.

For a realistic description of the splitting process we employ
MCTDHB(2) \cite{alon:08}, where the orbitals are
determined self-consistently from a variational principle. This
approach accounts in a natural manner for both the atom number
fluctuations and the orbital dynamics. We raise the issue, if the
nonadiabatic dynamics of the orbitals allows us to exploit the
same control mechanism as before.

\begin{figure}
  \centerline{\includegraphics[width=0.97\columnwidth]{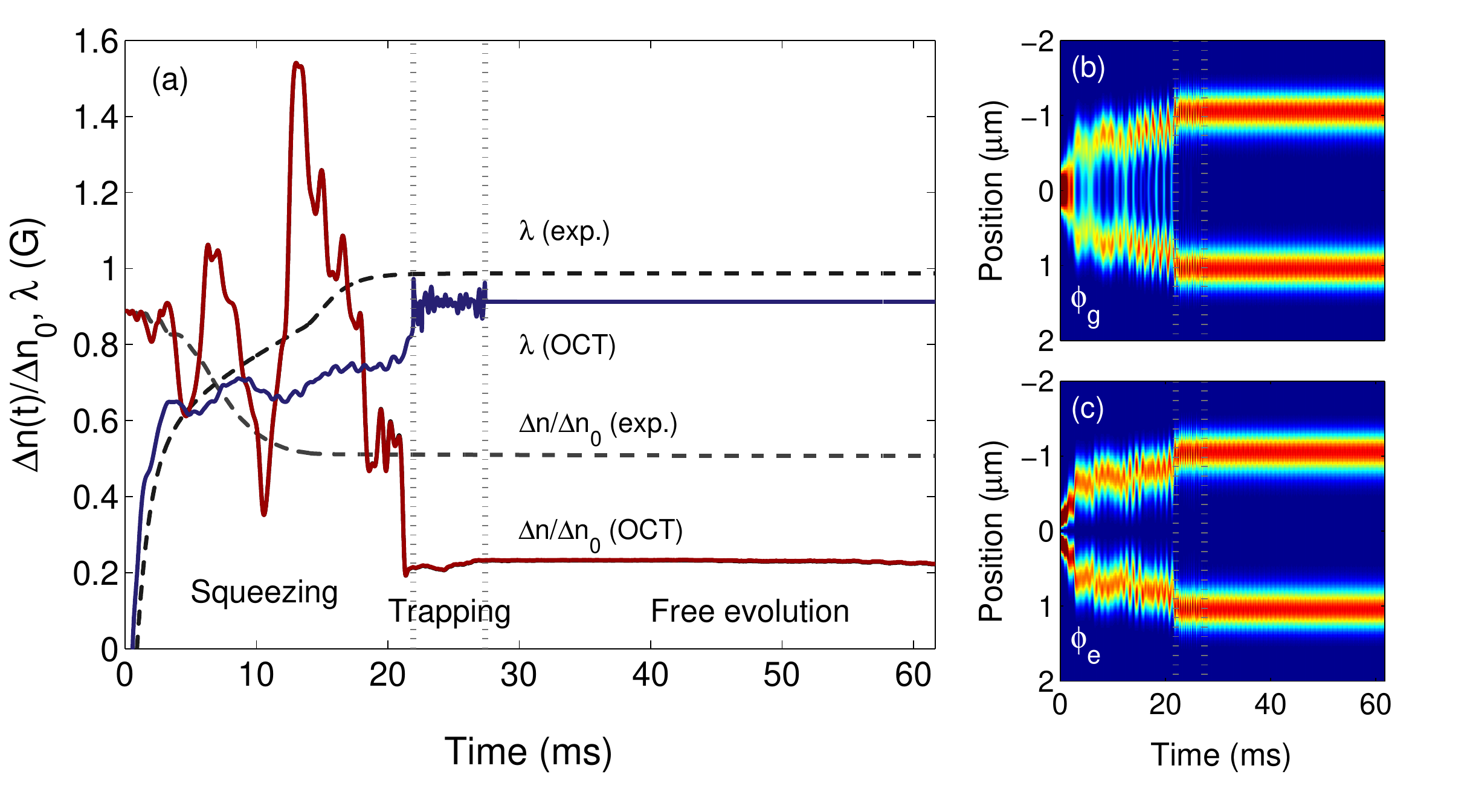}}
  \caption{(Color online) Optimized versus unoptimized splitting, as
  computed within MCTDHB(2). For the trap we use splitting by
  rf potentials \cite{lesanovsky:06}, with $\omega_T=(2\pi)2$
  kHz transversal frequency, and the control field $\lambda$ relates to 
  the amplitude of the rf field $B_{\rm rf}= (0.5+0.3\lambda)$ G.
  For an interaction strength \cite{alon:08} (chemical potential) of $g\approx 1.16$ Hz
  ($\mu\sim2$ kHz for the unsplit trap), OCT improves squeezing
  considerably as compared to the exponential case. In the
  calculations we use {\em gerade}\/ and {\em ungerade}\/ orbitals
  $\phi_{g,e}$, rather than $\phi_{L,R}$, as they allow to fully
  exploit the symmetry of the confinement potential. 
  The scheme is quite robust to typical experimental noise of $30\,\mu$G (black
  lines, indistinguishable from OCT results). Panel (a) reports the
  control fields and atom number fluctuations for $N=100$ for the
  exponential (dashed lines) and optimal control (solid lines), and
  panels (b) and (c) the time evolution of the {\em gerade}\/ and
  {\em ungerade}\/ orbitals. }
\end{figure}

In our calculations we use parameters typical for
cigar-shaped potentials with a few 100-500 atoms, split along
the transverse direction, similar to recent atom chip
\cite{schumm:05,hofferberth:07} or squeezing experiments
\cite{esteve:08} (see caption of Fig.~3 for details). We first
choose $\lambda(t)$ such that the tunneling decays approximately
monoexponentially. Figure~3 shows that for such control the number fluctuations decay in a fashion similar to the generic two-mode  model and the number squeezing in the final state is rather low. 

Squeezing optimization is again performed within the framework of
OCT \cite{remark.oct}. Fig.~3 shows the details of our OCT calculations. The density
plot of the orbitals clearly shows that, as in the two-mode model
calculation, the condensate is first brought to oscillations
within the two wells, resulting in an oscillating tunnel
coupling. In this regime the atom number fluctuations first
oscillate wildly and then significantly drop. These results are in good agreement with the generic two-mode model; see Fig.~1. 

To turn off the condensate oscillations after the squeezing
optimization, we introduced an additional optimization step for
the trapping of the orbitals, similarily to our previous work on
the optimization of the Gross-Pitaevskii equation
\cite{hohenester.pra:07}. With this the orbitals are brought to an
almost complete halt, as evidenced by the stationary evolution at
later times \cite{remark2.oct}.

We obtain similar results for larger atom numbers
(500 atoms) and different splitting times. This makes us believe
that the oscillating control for achieving a high degree of
squeezing is of general nature and that the simple two-mode model
provides a proper description for the underlying physics.


\begin{figure}
   \centerline{\includegraphics[width=0.9\columnwidth]{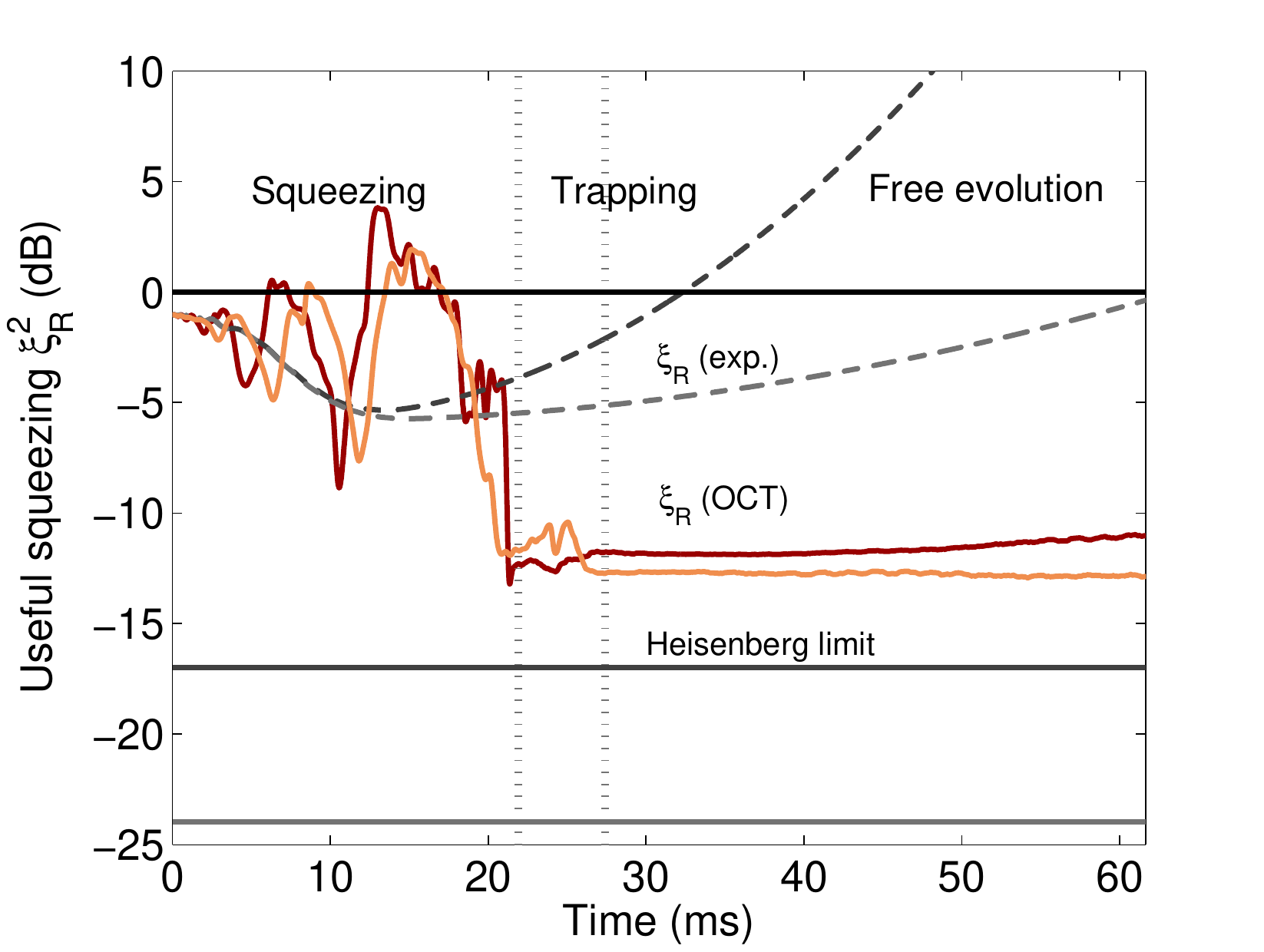}}
   \caption{(Color online) Useful squeezing $\xi_R$ for
   $N=100$ (dark lines) and $N=500$ (bright lines) for the
   exponential (dashed lines) and optimal control (solid lines). The
   optimized solutions stay well beyond the standard quantum limit ($\xi_R=1$), in sharp contrast to the unoptimized cases.
   The lines in the lower part indicate the Heisenberg limits $\xi_R=\sqrt{2/N}$ for the investigated $N$'s.}
\end{figure}

In the free-time evolution after splitting, an atom number superposition
state with finite $\Delta n$ undergoes a spread of evolution rates due to the nonlinear atom-atom interactions \cite{javanainen:97}. This ``diffusion''  of the relative phase with time degrades the coherence $\alpha:=2Re\langle \hat{a}_R^{\dagger}\hat{a}_L\rangle/N$ of the condensates, directly observable as the fringe contrast \cite{pitaevskii:01}. Here, $\hat{a}_L$ ($\hat{a}_R$) represents the mode operator for the left (right) condensate. To quantitatively analyze the improvement of OCT, we calculated the coherence $30$ ms after splitting for $N=100$ ($500$) atoms, and found values of above $80\%$ ($90\%$) for OCT, to be contrasted with the values of close to zero ($50\%$) for exponential splitting. {\em Thus our squeezing protocol strongly improves the phase coherence for a long time after splitting.}

Our OCT protocols will also be useful in experiments based upon the measurement of an atom number difference between the wells. The phase sensitivity is then quantified by the factor of {\em useful squeezing}\/ $\xi_R=2\Delta n/\sqrt{N}\alpha$ \cite{wineland:94}. It displays that squeezing enhances the sensitivity below the shot-noise limit $\xi_R=1$ and is (Heisenberg) limited from below by $\sqrt{2/N}$ \cite{giovannetti:04}. Furthermore, $\xi_R<1$ is a sufficient criterion for the presence of entanglement between the N atoms, signifying a type of entanglement which is useful as a resource \cite{esteve:08,sorensen:01}. It is apparent from Fig.~4 that OCT squeezing achieves a phase sensitivity close to this fundamental limit of quantum measurements as well as the creation of N-body nonseparable states.

In conclusion, we have demonstrated that non-adiabatic condensate splitting following OCT allows for a very efficient number squeezing on short time scales leading to a strongly enhanced phase coherence, thus rendering the technique powerful for interferometry applications.

We thank Alfio Borz\`\i, Greg von Winckel, Ofir Alon, and Thorsten
Schumm for most helpful discussions. This work has been supported
in part by the Austrian Science Fund FWF under project
P18136--N13.


\begin{thebibliography}{31}
\bibitem[{rem({\natexlab{a}})}]{remark.oct}
\bibinfo{note}{We use an analogous cost functional as in the case of the simple
  two-mode model [Eq.~\eqref{eq:cost}], but with the system constrained to the
  time evolution of the MCTDHB(2) equations. This results in a much more
  involved optimality system as compared to the former model. Details of the
  optimality system and our numerical implementation will be given elsewhere.}
\bibitem[{rem({\natexlab{b}})}]{remark2.oct}
\bibinfo{note}{Also in the stage termed 'squeezing' optimization, we used a
  cost functional comprised of both the number fluctuations at the final time
  and terms accounting for the trapping of the orbitals.}
\end{thebibliography}

\end{document}